\documentclass{article}

\usepackage{microtype}
\usepackage{graphicx}
\usepackage{subcaption}
\usepackage{booktabs} 

\usepackage{hyperref}

\usepackage[accepted]{icml2025}

\usepackage{amsmath}
\usepackage{amssymb}
\usepackage{mathtools}
\usepackage{amsthm}
\usepackage{subcaption}

\usepackage[capitalize,noabbrev]{cleveref}

\theoremstyle{plain}

\theoremstyle{definition}

\theoremstyle{remark}

\usepackage[textsize=tiny]{todonotes}

\icmltitlerunning{Isotropy and Geometry of Pretrained Protein LMs}

\begin{document}

\twocolumn[
\icmltitle{A Look at the Isotropy of Pretrained Protein Language Models}



\icmlsetsymbol{equal}{*}

\begin{icmlauthorlist}
\icmlauthor{Sheikh Azizul Hakim}{equal,yyy}
\icmlauthor{Kowshic Roy}{equal,yyy}
\icmlauthor{M Saifur Rahman}{yyy}

\end{icmlauthorlist}

\icmlaffiliation{yyy}{Department of Computer Science and Engineering, Bangladesh University of Engineering and Technology, Dhaka, Bangladesh}

\icmlcorrespondingauthor{M Saifur Rahman}{mrahman@cse.buet.ac.bd}

\icmlkeywords{Machine Learning, ICML}

\vskip 0.3in
]



\printAffiliationsAndNotice{} 




\begin{abstract}
Large pretrained language models have transformed natural language processing, and their adaptation to protein sequences---viewed as strings of amino acid characters---has advanced protein analysis. However, the distinct properties of proteins, such as variable sequence lengths and lack of word-sentence analogs, necessitate a deeper understanding of protein language models (LMs). We investigate the isotropy of protein LM embedding spaces using average pairwise cosine similarity and the IsoScore method, revealing that models like ProtBERT and ProtXLNet are highly anisotropic, utilizing only 2--14 dimensions for global and local representations. In contrast, multi-modal training in ProteinBERT, which integrates sequence and gene ontology data, enhances isotropy, suggesting that diverse biological inputs improve representational efficiency. We also find that embedding distances weakly correlate with alignment-based similarity scores, particularly at low similarity. 
\end{abstract}

\section{Introduction}

The most sophisticated machines in nature are proteins. Across the tree of life, proteins play a crucial role in catalyzing biochemical reactions, providing structural support for other cell organelles, facilitating cell signaling, contributing to immune defense, and even synthesizing other proteins \cite{Kessel2018Mar}. Proteins can be regarded as sequences of amino acid characters. Consequently, machine-learning techniques tailored for natural language and other sequences are ideally suited for forecasting protein-related tasks \cite{Ofer2021Jan}. 

In recent years, natural language processing has been significantly advanced with the advent of large pretrained attention-based language models such as BERT \cite{BERT}, XLNet \cite{xlnet}, Albert \cite{albert}, GPT \cite{Radford2019} etc. (for a detailed survey, check \cite{Min2023Sep}). The same concepts, and often the same architectures, have been applied to proteins \cite{Rao2019,prottrans,pbert}. The attention mechanism, in particular, has been shown to correlate with many known biological and biochemical properties \cite{Vig2020Jun}. However, prior works have also noted that protein sequences behave differently from natural languages; for example, protein sequences can vary significantly in length, from under fifty amino acids to over thousands, unlike words and sentences, and we cannot break down proteins into analogs of words and sentences in the first place \cite{pbert}. 

Language models (LM) use several types of embeddings that map a linguistic concept into a geometric space. Traditionally, static embeddings have been utilized \cite{glove}, and such approaches have been theoretically explained as the factorization of a word-context matrix containing a co-occurrence statistic \cite{Levy2014Jan, Levy2014Dec}. Theoretical and empirical evidence suggests that many of these models are \textit{isotropic}, i.e., angularly uniform \cite{Arora2016Dec}.  However, context-sensitive word representations can be found from pretrained language models, such as BERT \cite{BERT}, GPT \cite{Radford2019}, and are useful for several downstream tasks. Ethayarajh \cite{ethayarajh} investigated the isotropic properties of the contextualized embedding spaces of such pretrained models using average pairwise cosine similarity. The cosine similarity of two vectors $\mathbf{x}$ and $\mathbf{y}$ is defined as the normalized dot product between them $\left( \frac{\mathbf{x.y}}{\mathbf{|x||y|}} \right )$. The contextual embedding spaces of the pretrained LMs came out to be, somewhat surprisingly, \emph{highly anisotropic}. Increasing isotropy has been suggested as a way to improve the performance of BERT \cite{bertology}, but \cite{Rajaee2021} showed that increasing isotropy using existing methods of post-processing pretrained LMs may hurt performance. \cite{cai2021isotropy} argued that a different notion of isotropy might indeed exist for the contextualized embedding spaces and identified some other geometric properties.

\cite{isoscore} argued that all existing measures of measuring isotropy have fundamental shortcomings. They identified some key properties of isotropy, such as mean agnosticity, global stability, rotational invariance, etc., and proposed a new scoring method, named \emph{IsoScore}, based on the covariance matrix of the principal components. They also showed that this score can be used to approximate the number of dimensions effectively used by the point cloud in consideration.

Although several works have been done to analyze the isotropy and geometry of the embedding spaces for natural languages, the attempt to do so is scarce (if any) for protein sequences. We analyze both the cosine similarity-based and IsoScore-based approaches to analyze the isotropy of protein embedding spaces. We find that protein LMs are highly anisotropic, and a much lower dimensional embedding space might come equally handy for downstream tasks. 
We also find that protein embedding distances (cosine and Euclidean) exhibit weak overall correlations with traditional alignment-based similarity scores, reliably capturing biological relationships only at high similarity; at low similarity, their high variance highlights limitations in representing distant relationships, underscoring the need for multi-modal models to integrate diverse biological signals.

To extend our analysis, we investigate the isotropy and geometry of local (per-residue) embeddings in protein language models, finding them to be highly anisotropic, utilizing only approximately 14 dimensions on average across models (Table 3). By visualizing these embeddings in a 3D space defined by the first three principal components, we observe distinct clustering patterns for each amino acid, suggesting that local embeddings capture residue-specific biochemical properties. These findings indicate significant redundancy in local representations, similar to global embeddings, and highlight opportunities for dimensionality reduction in multi-modal protein models.


The contribution of this study is to explore various properties of protein embedding spaces. At first, We find that protein LMs are highly anisotropic, and a much lower dimensional embedding space might come equally handy for downstream tasks. Then, we explore the relationship between distances in embedding space and the alignment distances between the protein sequences. We extend the same result of anisotropy in the case of local (per-residue) representations. We also explore the geometry of the local embeddings for each amino acid. (\#TODO: rewrite the paragraph)

\section{Materials and Methods}

\subsection{Dataset}

We use the SwissProt subset of the UniProt database \cite{uniprot2023uniprot}, consisting of approximately 570,000 protein sequences with experimentally validated annotations. SwissProt is manually curated and includes high-quality functional and structural information, making it a reliable benchmark for evaluating protein language models. Its focus on experimentally verified proteins ensures that downstream tasks—such as similarity analysis or embedding evaluation—are grounded in biologically meaningful data.

\subsection{Protein Language Models in Consideration}

We evaluated three pretrained protein language models from \cite{prottrans}: ProtXLNet, ProtBERT, and ProtBERT-BFD, the latter trained on a distinct dataset. Pretrained weights were obtained from Hugging Face\footnote{\url{https://huggingface.co/Rostlab}}. In these models, protein sequences are treated as sentence-like sequences, with each amino acid residue represented as a word-like token. The underlying architectures, adapted from their natural language counterparts (XLNet and BERT), remain unmodified. These models generate per-residue (local) embeddings for input proteins, with per-protein (global) embeddings derived through average pooling of local embeddings. \cite{prottrans} explored alternative pooling strategies, including minimum, maximum, and concatenation pooling, but found average pooling to be the most effective for generating robust global representations. 

We also evaluated ProteinBERT from \cite{pbert}, which employs a distinct architecture tailored for protein modeling. Unlike sequence-only models, ProteinBERT is trained on both protein sequences and gene ontology (GO) annotations, enabling a multi-modal approach that captures functional and structural insights. Its architecture directly generates both per-residue (local) and per-protein (global) embeddings, eliminating the need for pooling local embeddings to derive global representations. Pretrained weights were obtained from the model’s GitHub repository\footnote{\url{https://github.com/nadavbra/protein_bert}}.

\section{Results and Discussion}

\subsection{Anisotropy of Global Embeddings}

\begin{table*}
    \centering
    \begin{tabular}{lccc}
        \hline
        \textbf{Model Name} & \textbf{Embedding Dimension} & \textbf{IsoScore} & \textbf{Effectively Used Dimensions} \\ \hline
        ProtBERT & 1024 & 0.001658146 & 3\\
        ProtBERT-BFD & 1024 & 0.003967522 & 6\\
        ProtXLNet & 1024 & 0.001502474 & 3\\
        ProteinBERT & 512 & \textbf{0.231227934} & \textbf{120}\\ 
        \hline
    \end{tabular}
    \caption{Embedding dimension, IsoScore, and effectively used dimensions for different protein language models.}
    \label{tab:iso_table}
\end{table*}

\begin{table*}[h]
    \centering
    \begin{tabular}{lrrrr}
    \hline
    Model Name & \textbf{cosine} & \textbf{sq\_euclidean} & \textbf{alignment\_score} & \textbf{similarity\_score} \\
    \hline
    \textbf{cosine} & 1.000000 & 0.791068 & 0.013804 & -0.011159 \\  
    \textbf{sq\_euclidean} &  & 1.000000 & -0.102698 & -0.145814 \\ 
    \textbf{alignment\_score} &  &  & 1.000000 & 0.847258 \\ 
    \textbf{similarity\_score} &  &  &  & 1.000000 \\ 
    \hline
    \end{tabular}
    \caption{Correlation matrix between different distance metrics for ProtBERT.}
    \label{tab:correlation_metrics}
\end{table*}

We computed IsoScores for embeddings generated by protein language models (LMs). This metric quantifies isotropy through mean agnosticity, global stability, and rotational invariance \cite{isoscore}. If the IsoScore of a pointcloud $\left(\mathbf{X} \in \mathcal{R}^n\right)$ is $i(\mathbf{X})$, then according to \cite{isoscore}, effectively $\dim(\mathbf{X}) = \left(i(\mathbf{X}) \times (n - 1) + 1\right)$ dimensions are utilized. If $i(\mathbf{X}) \approx 0$, then the pointcloud is highly anisotropic, and no more than one dimension is being effectively utilized. If $i(\mathbf{X}) \approx 1$, then the pointcloud is highly isotropic and all the dimensions are being effectively utilized. We tabulate the IsoScores and the number of effectively utilized dimensions (fractions are rounded up to the next integer) in Table \ref{tab:iso_table}. We find that the protein LMs developed \cite{prottrans}, which is trained using the eponymous model architectures built for natural language, under consideration are \emph{highly anisotropic} and use very few (2-5) dimensions. On the other hand, ProteinBERT has a relatively high isotropy score (0.23) and uses 120 dimensions effectively. For comparison purposes, the reported IsoScores in \cite{isoscore} for BERT and GPT  are 0.11 and 0.18, respectively. Thus, while protein LMs using traditional architectures are, in general, more anisotropic than natural LMs, ProteinBERT is more isotropic. We think this is because ProteinBERT uses a different architecture and is trained not only from protein sequences, but also from gene ontology (GO) annotations. ProteinBERT's architecture enables it to output global and local representations separately, instead of the other models, where local embeddings are pooled to generate global embeddings.


\subsection{Comparison between Alignment Distances and Embedding Distances}

We investigated the relationship between traditional similarity measures (alignment score and similarity score) and embedding-based measures (squared Euclidean distance and cosine similarity). We used BioPython \cite{cock2009biopython} to calculate the alignments using the PAM-250 scoring matrix \cite{dayhoff1978model}. We define similarity score as the fraction of identical residues in optimal alignments. For this experiment, we randomly sampled 1\% of the Swissprot proteins which resulted in $6.4\times10^6$ protein pairs. Our results show that the two traditional measures are strongly correlated with each other, as are the two embedding-based measures. However, correlations between traditional and embedding-based metrics are weaker—often low or even negative—suggesting that these approaches do not capture the same aspects of protein similarity. We report the pairwise correlation coefficients of the four similarity measures, as calculated for ProtBERT, in Table \ref{tab:correlation_metrics}. Other models exhibit similar trends.   

\begin{figure*}[h]
    \centering
    \begin{subfigure}{0.49\linewidth}
        \centering
        \includegraphics[width=\linewidth]{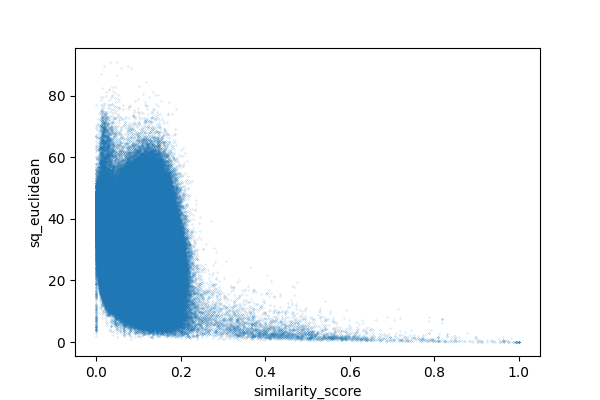}
        \caption{Squared Euclidean distance of ProteinBERT embeddings vs alignment similarity score.}
        \label{fig:sqeucnlid_simscore_ProtBERT}
    \end{subfigure}
    \hfill
    \begin{subfigure}{0.49\linewidth}
        \centering
        \includegraphics[width=\linewidth]{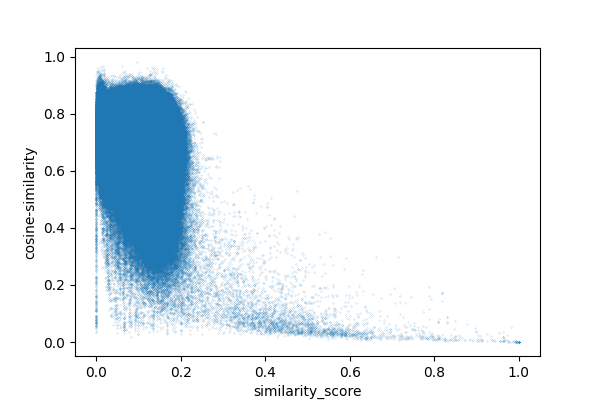}
        \caption{Cosine similarity of ProteinBERT embeddings vs alignment similarity score.}
        \label{fig:cosine_simscore}
    \end{subfigure}
    \caption{Relationship between embedding-based distances and traditional alignment-based similarity scores.}
    \label{fig:embedding_vs_alignment}
\end{figure*}

We further investigated the relationship between embedding-based distances and traditional alignment-based similarity scores and observed consistent non-linear patterns. As shown in Figures \ref{fig:sqeucnlid_simscore_ProtBERT} and \ref{fig:cosine_simscore}, while the overall correlations are weakly negative, a clear structural trend emerges: for low similarity scores, both squared Euclidean and cosine distances exhibit high variance and span a wide range of values, indicating poor predictive power. In contrast, at high similarity scores, both metrics converge—Euclidean distances become consistently low, and cosine similarities cluster near 1.0. This asymmetric behavior suggests that embedding distances, while informative at the high end of biological similarity, are unreliable indicators of similarity in the low-alignment regime, revealing a key limitation in how current embeddings capture biologically meaningful relationships. While Figures \ref{fig:sqeucnlid_simscore_ProtBERT} and \ref{fig:cosine_simscore} are generated for ProteinBERT, this trend generalizes across all models in consideration. 

\subsection{Anisotropy of Local Embeddings}

To investigate the geometric structure of protein embedding spaces, we measured the IsoScore of individual amino acid token embeddings. Table~\ref{tab:amino_model_values} reveals a clear anisotropy in the embedding space for each amino acid.  For 1024 embedding dimensions for each of the three models, only about 14 dimensions are effectively used on average.



\section{Conclusion}


Our analysis reveals a critical mismatch between biological richness and embedding geometry in pretrained protein language models. Sequence-only models like ProtBERT and ProtXLNet produce highly anisotropic embeddings that utilize minimal representational capacity, while multi-modal ProteinBERT demonstrates improved isotropy through biological priors. 

These results have direct implications for generative biology, where diverse and informative latent spaces are essential for tasks such as protein design, variant prediction, and molecular optimization. The strong anisotropy in current embeddings suggests that models may fail to explore biologically meaningful subspaces during generation, leading to reduced diversity or biological invalidity. Moreover, we observe that learned distances in embedding space—based on cosine and Euclidean metrics—correlate poorly with biologically grounded similarity measures like sequence alignment, particularly at low similarity. This divergence implies that embedding spaces lack robustness when modeling distant or novel proteins, a serious limitation for generative models that seek to extrapolate beyond known data.

In a concurrent work, \cite{10.1093/bib/bbaf047} analyzed the phylogenetic properties captured by protein LMs. Future studies might look for if improving isotropy leads to better phylogenetic relationships.

Looking ahead, we advocate for the development of next-generation protein LMs that explicitly optimize for geometric richness, isotropy, and biological alignment. Promising directions include biologically supervised contrastive pretraining, isotropy-promoting regularization, and functional embedding constraints grounded in ontologies or structural data. Such efforts could produce embeddings that are simultaneously compact, generative, and biologically meaningful—making them ideal backbones for AI-driven discovery in protein science. By better understanding and shaping the geometry of protein embedding spaces, we lay the groundwork for interpretable, multi-modal, and experimentally actionable generative models in biology.

\subsection*{Code Availability:} Our implementation and the generated plots can be found in \url{https://github.com/vodro/geometry_of_proteins}.

\begin{table}[H]
\centering
\begin{tabular}{lrrr}
\toprule
\textbf{Amino Acid} & \textbf{BERT} & \textbf{BERT-BFD} & \textbf{XLNet} \\
\midrule
Alanine (A)     & 0.013340 & 0.017366 & 0.011098 \\
Cysteine (C)    & 0.012388 & 0.013517 & 0.010462 \\
Aspartic Acid (D) & 0.012656 & 0.013981 & 0.011726 \\
Glutamic Acid (E) & 0.013102 & 0.017743 & 0.012512 \\
Phenylalanine (F) & 0.013049 & 0.012437 & 0.010798 \\
Glycine (G)     & 0.011422 & 0.011228 & 0.009837 \\
Histidine (H)   & 0.011963 & 0.011971 & 0.011251 \\
Isoleucine (I)  & 0.013796 & 0.012815 & 0.011363 \\
Lysine (K)      & 0.012053 & 0.021384 & 0.012672 \\
Leucine (L)     & 0.013934 & 0.013625 & 0.011241 \\
Methionine (M)  & 0.015887 & 0.017329 & 0.010793 \\
Asparagine (N)  & 0.010028 & 0.017633 & 0.010436 \\
Proline (P)     & 0.011258 & 0.011756 & 0.011503 \\
Glutamine (Q)   & 0.012816 & 0.020504 & 0.011812 \\
Arginine (R)    & 0.012033 & 0.012910 & 0.012437 \\
Serine (S)      & 0.010018 & 0.017425 & 0.009935 \\
Threonine (T)   & 0.011633 & 0.014803 & 0.010239 \\
Valine (V)      & 0.014402 & 0.013638 & 0.010828 \\
Tryptophan (W)  & 0.012764 & 0.011212 & 0.011547 \\
Tyrosine (Y)    & 0.013153 & 0.012430 & 0.011601 \\
Unknown (X)     & 0.010520 & 0.006424 & 0.007258 \\
\bottomrule
\end{tabular}
\caption{Per-amino acid IsoScore values for three models: BERT, BERT-BFD, and XLNet, rounded to six decimal places.}
\label{tab:amino_model_values}
\end{table}

\pagebreak  
\bibliography{example_paper}
\bibliographystyle{icml2025}



\end{document}